\documentclass{PoS}
\usepackage{graphicx}
\usepackage{subfig}
\usepackage{amssymb}
\usepackage{amsmath}
\def\igrj{IGR J$17252-3616$}
\def\nh{N$_{\rm H}$}

\title{The nature of the absorber in the INTEGRAL highly obscured sgHMXB IGR J17252-3616}

\ShortTitle{The nature of the absorber in sgHMXB IGR J17252-3616}

\author{\speaker{Antonios Manousakis}%
\\
       ISDC Data Center for Astrophysics, Switzerland\\
       E-mail: \email{Antonios.Manousakis@unige.ch}}

\author{Roland Walter\\
        ISDC Data Center for Astrophysics, Switzerland\\
}

\abstract{INTEGRAL played a key role in discovering obscured sgHMXB in the Galaxy. We used XMM- Newton to perform X-ray wind tomography of a specific of these systems, IGR J17252-3616, featuring eclipses of the accreting pulsar. The X-ray band (0.2-10 keV) reveals vital information on the geometry of the surrounding gas probing simultaneously the absorption and the fluorescence emission. The XMM observations were scheduled to cover as many orbital phases as possible. Timing analysis allows the derivation of an accurate orbital solution and of the system parameters. Spectral analysis revealed remarkable variations of the absorbing column density along the orbit and of the Fe K$\alpha$ fluorescence line around the eclipse. The combination of these observables revealed a highly asymmetric and unprecedentedly extended structure in the stellar wind extending up to 2-3 stellar radii. The observations can be modeled in terms of three independent components: \\
i) the unperturbed stellar wind\\
ii) the contribution of a highly asymmetric hydrodynamic wind tail-like structure and\\
iii) a cusp of material close to the neutron star.\\
These dynamical structures are imaged for the first time in a sgHMXB and explain the source of the high obscuration.}

\FullConference{8th INTEGRAL Workshop `The Restless Gamma-ray Universe'\\
		 September 27-30 2010\\
		 Dublin Castle, Dublin, Ireland}

\begin{document}

\section{Introduction}
High mass X-ray binaries (HMXB) consist of a neutron star or a black hole fueled by the accretion of
the wind of an early-type stellar companion.
Their X-ray emission, a measure of the accretion rate, shows a variety of transient to persistent patterns.
Outbursts are observed on timescales from seconds to months and dynamical ranges varying by factors of $10^4$.
The majority of the known HMXB are Be/X-ray binaries, with Be stellar companions. These systems are transient, 
featuring bright outbursts with typical durations on the order of several weeks \cite{White1989}. 
A second class of HMXBs harbor OB supergiant companions (sgHMXBs) that feed the compact object 
by means of strong, radiatively driven stellar winds or Roche lobe overflow. 
Thanks to INTEGRAL, the number of known sgHMXB systems has tripled in the past few years \cite{Walter_et_al06}.

\igrj\, was detected by $ISGRI$ onboard $INTEGRAL$ on February 9, 2004 among other hard X-ray sources \cite{Walter}. 
The source was first detected by $EXOSAT$ (EXO 1722-3616) as a weak soft X-ray source, back in 1984 \cite{Warwick}. 
In 1987, $Ginga$ performed a pointed observations and revealed a highly variable X-ray source, X1722-363, with a pulsation period  
of $\sim 413.9$ sec \cite{Tawara}. Additional $Ginga$ observations revealed the orbital period of 9-10 days and a
mass of the companion star of $\sim$ 15 $M_{\odot}$ \cite{Takeuchi}. Both studies concluded that the system is a high mass X-ray binary (HMXB).

$INTEGRAL$ and $XMM-Newton$ observations of \igrj\, allowed Zurita-Heras et al. (2006) to identify the infrared counterpart 
of the system, to accurately measure the absorbing column density, 
and refine the spin period of the system \cite{Zurita}. Thanks to the eclipses, an accurate orbital period could be derived from $INTEGRAL$ data. 
Further RXTE observations helped identify a highly inclined system ($i>61^{o}$) with a companion star of $\rm{M}_{*}\lesssim20\, \rm{M}_{\odot}$ and $\rm{R}_{*}\sim 20-40 \,\rm{R}_{\odot}$ \cite{Thompson, Corbet}.
Recent VLT observations help to infer the companion spectral type (between B0--B5 I and B0--B1 Ia) and radial velocity measurements \cite{Mason}. 
Its spectral energy distribution  can be characterized by a temperature of $\rm{T}_{*}\sim 30$ kK and a reddening of  A$_{V}\sim 20$ \cite{Rahoui}.

\section{Results}

The Science Analysis Software (XMM-SAS)  version 
9.0.0\footnote{http://xmm.esac.esa.int/sas/}  was used  to produce event lists for the EPIC-pn instrument. 
Spectra and lightcurves were built by collecting double and single events in the energy range 0.2 - 10 keV. The observations were scheduled 
to cover the orbital phases, 0.01, 0.03, 0.08, 0.15, 0.27, 0.37, 0.40, 0.65, 0.79, and 0.91 inferred from the orbital solutions.

\subsubsection*{Orbital solution}
We have obtained a refined orbital solution based on RXTE \cite{Thompson} and XMM-Newton Pulse Arrival Times  \cite{Manousakis} to derive accurate orbital phases.
The orbital period is constrained using INTEGRAL\footnote{HEAVENS \cite{HEAVENS} interface available at www.isdc.unige.ch/heavens} long-term monitoring. 
We have obtained  an orbital period of $P=9.742 \pm 0.001$ days, a projected semi-major axis $\alpha_{\chi}$sin$i$= 102 $\pm$ 8 lt-s, a mid-eclipse $T_{90} = MJD 53761.67 \pm 0.1$ and an 90\% 
upper limit $e<0.15$ for eccentricity, yielding a mass function $f\sim 11.7$ M$_{\odot}$. 
Corresponding masses of both donor star and compact object are in the range, $M_{OB}\approx$ 14 -- 17 M$_{\odot}$
 and $M_{NS}\approx$1.4  -- 1.7 M$_{\odot}$, respectively.  

\subsubsection*{Spectral Variability}
The spectra are always heavily absorbed  below $\sim$ 3 keV and contain an iron K-edge at 7.2 keV. 
Throughout our observations  an iron K$\alpha$ line, at 6.40$\pm$0.03  keV, is always present.  
Some parameters (photon index, cut-off energy, blackbody templrature)  did not vary significantly among the observations.  
To search for spectral variability we decided to fix them to their averages values (E$_{C}$=8.2 keV, $\Gamma=0.02$, kT$_{BB}$=0.5 keV). 
The model was built by using  an intrinsically  absorbed high energy cut-off power-law with a gaussian line, 
a black-body component responsible for the observed soft X-ray excess,
and an overall  absorption  responsible for the interstellar absorption fixed to the Galactic absorption N$_{H}$=1.5$\times 10^{22}$ cm$^{-2}$. The model can be described as  \verb=wabs*(bbody + gauss + vphabs*cutoff)=
in XSPEC\footnote{heasarc.gsfc.nasa.gov/docs/xanadu/xspec/} using 
standard abundances obtained from \cite{abundance}.
All the normalization parameters and the intrinsic  column density were free to vary. 
Figure 1 shows the variations of intrinsic N$_{H}$ (left) and the iron K$\alpha$ line corrected equivalent width\footnote{
As the X-ray continuum illuminating the gas emitting the Fe fluorescent line cannot be measured during the eclipse, we calculated 
a corrected Fe K$\alpha$ EW by assuming a constant continuum flux of $1.8\times 10^{-3}$ ph\, keV$^{-1}$ cm$^{-2}$ s$^{-1}$} (right) as 
a function of phase together with a wind model described in the next section. Huge variations of $N_{H}$ can be seen, scaling up to a factor of 10. 
The unabsorbed 0.2-10 keV luminosity level (outside the eclipse)  is compatible with $L_{X}\sim 10^{36}$ erg s$^{-1}$, assuming a distance of 8 kpc \cite{Mason}.
Minor variations on the flux can be interpreted as variation of the instantaneous accretion rate ($\dot{M}$) onto the neutron star. Inside the eclipse the flux drops by a factor 
of $\sim$ 200. In the following analysis, we assumed a distance $D\approx$ 8 kpc, a circular orbit ($e=0$), and an edge-on geometry ($i=90^\circ$).

\begin{figure}
  \centering
  \includegraphics[height=0.40\textwidth,width= 0.38\textwidth,angle=0]{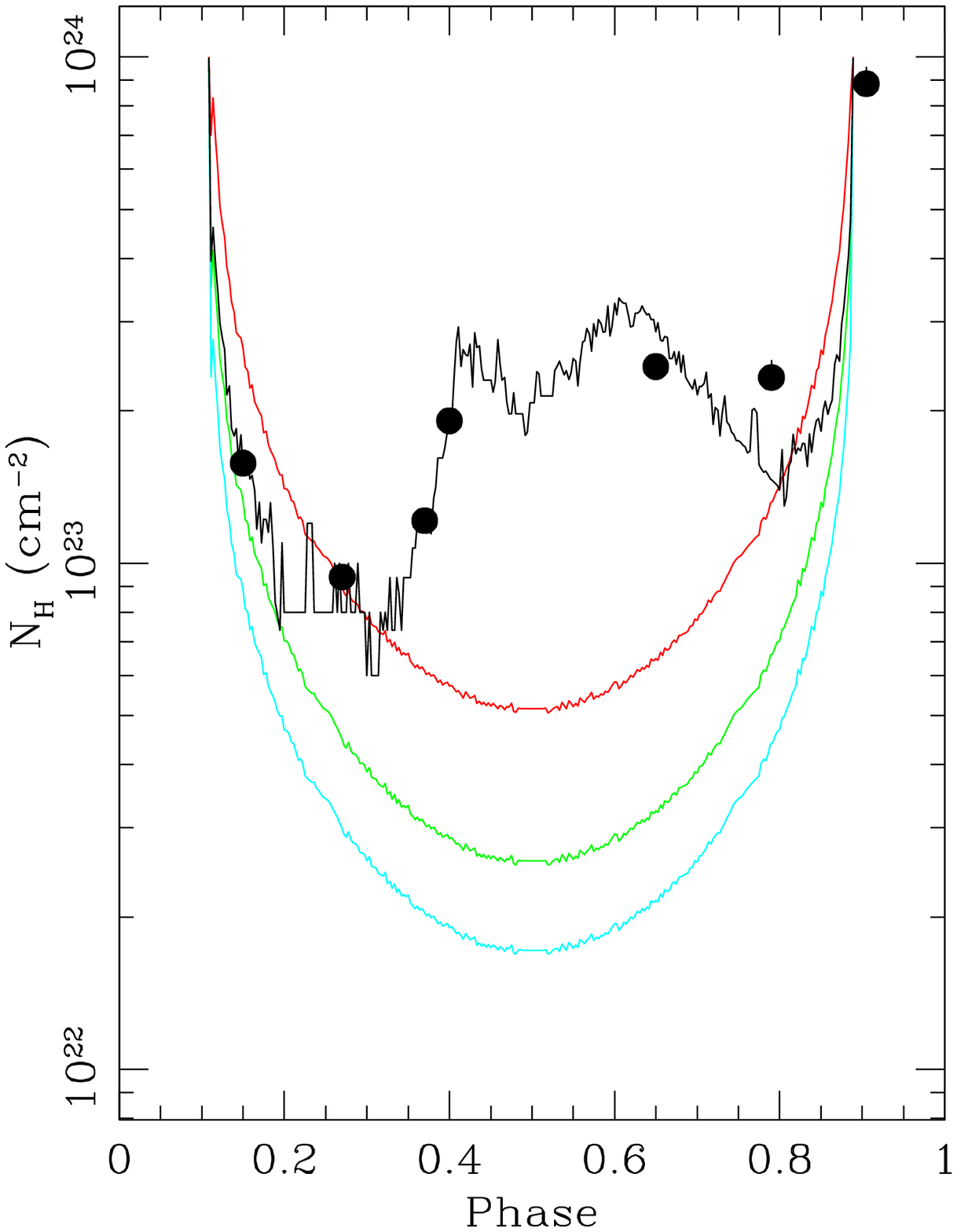} \hspace{1.5cm}
     \includegraphics[height=0.40\textwidth,width= 0.38\textwidth,angle=0]{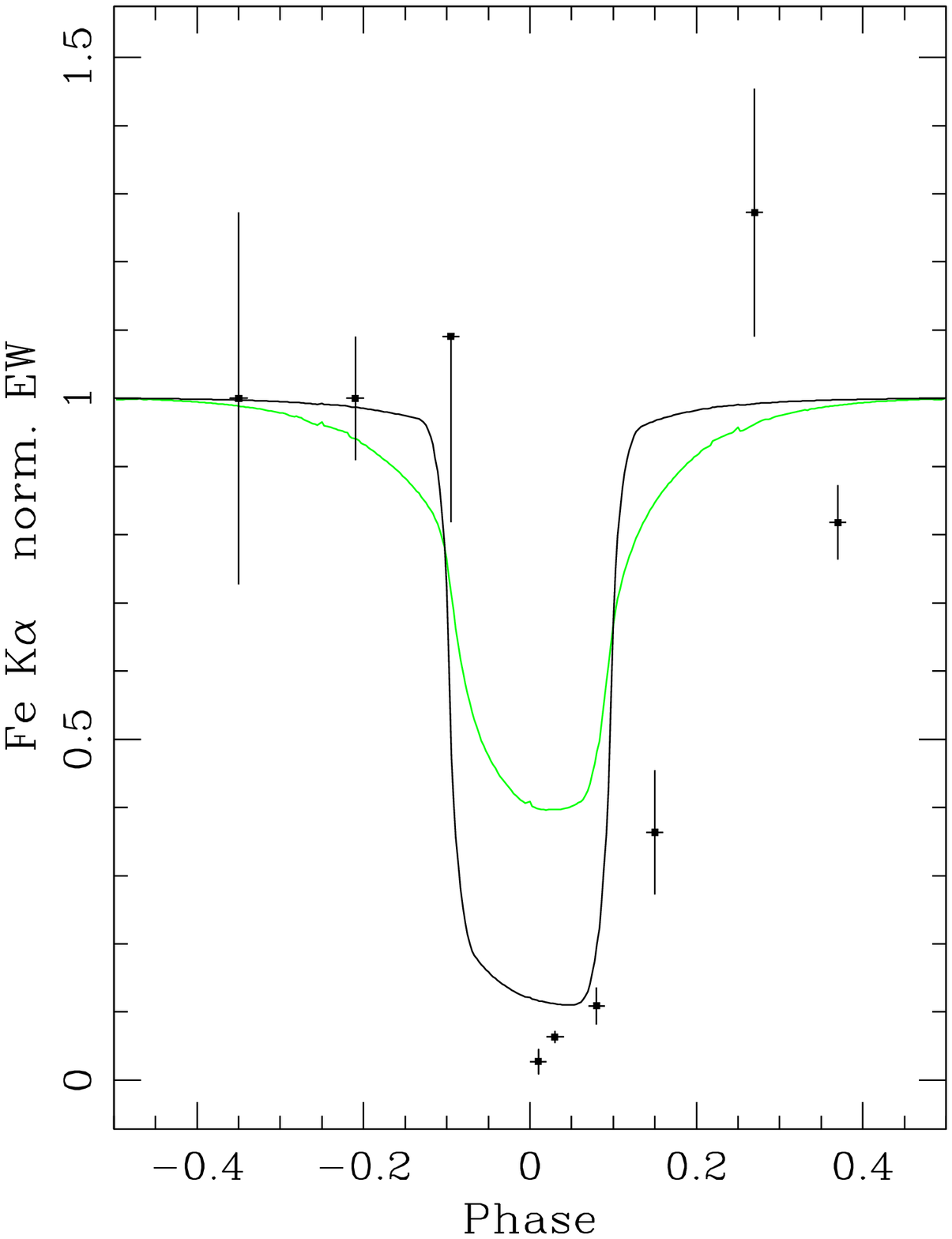}
  \caption{\emph{Left panel:} Simulated N$_{H}$ profiles together with the data. 
 Smooth stellar wind is illustrated by cyan, green, and, red curves as a function of $\dot{M}/\upsilon_{\infty}$=0.7, 1, 2$\times 10^{-16}$ M$_{\odot}$/km, respectively. 
 Black solid line shows the total N$_{H}$ consisting of the unperturbed stellar wind (green curve) and the tail-like extended component. 
  \emph{Right panel:} Iron K$\alpha$ corrected equivalent width during the orbit. Green curve indicated the perdition of the hydrodynamical 
  tail and the black curve the addition of the central cocoon. }
  \label{fig:animals}
\end{figure}

\section{Discussion}

\subsubsection*{Variability of absorption along the orbit}   \label{text:nhprofile}

We constructed \cite{Manousakis} a 3D model of the OB supergiant stellar wind to
investigated the behavior of the intrinsic column density, \nh, as a function of phase, and to identify potential structure of the wind during the orbit.  
We approximated the wind structure with two components, the unperturbed wind ($\rho_{wind}$) and a tail-like hydrodynamic perturbation ($\rho_{tail}$) 
related to the presence of the neutron star. 
 These shocks are produced by 
 hydrodynamical  simulations \cite{Blondin} but produce a \nh\, of up to 
$\sim$10$^{22}$ cm$^{-2}$, which is too small to account for the variability observed in IGR J17252--3616. 

The unperturbed stellar wind was modeled  by assuming a
standard  wind profile \cite{CAKwind}
\begin{displaymath}
\upsilon(r)=\upsilon_{\infty}\Big(1-\frac{R_{*}}{r}\Big)^{\beta},
\end{displaymath}
where $\upsilon(r)$ is the wind velocity at distance $r$ from the stellar center, 
$\upsilon_{\infty}$ is the terminal velocity of the wind, and $\beta$ is a parameter describing the wind gradient. 
The conservation of mass provides the radial density distribution of the stellar wind.
The unperturbed stellar wind is a good approximation within the orbit of the neutron star.
Hydrodynamical simulations \cite{Blondin} of HMXB  have shown that the wind can be highly disrupted by the neutron star
beyond the orbit.

To estimate the terminal velocity of the unperturbed wind, 
we studied the \nh\ variability using three different sets of parameters (Fig. 1 left).
The mass-loss rate and terminal velocity are constrained by the data to be in the range
$\dot{M}_{w}/\upsilon_{\infty}\sim (0.7-2)\times10^{-16}$ M$_{\odot}$/ km ($\beta$ has a very limited impact on the results, so  we used 0.7).

The fraction of the wind captured by the neutron star could be estimated from the accretion radius $r_{acc}= 2 GM_{X}/(\upsilon_{orb}^2+\upsilon^2)\sim 2\cdot 10^{11} ~{\rm cm}$ 
(where $\upsilon_{orb}=250~{\rm km/s}$ is the orbital velocity) as $f \sim \pi r_{acc}^{2} / 4\pi R_{orb}^{2} \sim 7.5\cdot 10^{-4}$. 
The mass-loss rate is therefore $\dot{M}_{w}\sim f^{-1}\dot{M}\approx 1.5\cdot10^{-6}$ M$_{\odot}$/yr and the terminal velocity of the wind
is constrained to be in the interval $\upsilon_{\infty}\sim$ 250 -- 600 km/h. 

In our simulation, we adopted a terminal velocity $\upsilon_{\infty}$=400 km/sec, a stellar radius $R_{*}$=29 $R_{\odot}$, 
a  wind gradient $\beta$=0.7, and a mass loss rate $\dot{M_{*}}$= 1.35$\times10^{-6}$ M$_{\odot}\,$ yr$^{-1}$.

We assumed that 
the tail-like structure is created very close to the neutron star and opens up with distance.
The density of the material inside the `tail' decreases with distance to ensure  mass conservation.  
Its distribution follows a `horn'-like shape with a  circular section.
We adjusted the density of the tail-like structure to match the observations.
The density distribution $\rho_{wind}+\rho_{tail}$ is displayed in
figure 2 left panel. 
The supergiant is located at the center (black disk). The tail-like structure covers about half  of the orbit.

Figure 1 (left panel) displays 
the simulated \nh\, variability from the above density distribution together with the observed data points. 
The data and the model shows that the tail-like perturbation is essential to understand the observed variations.

\subsubsection*{Variability of the Fe K$\alpha$ line during the orbit} \label{text:feprofile}

Assuming that the intrinsic X-ray flux is unaffected by the eclipse, the Fe K$\alpha$ equivalent width drops by 
a factor $\sim 10$ during the eclipse in an orbital phase interval of $\sim 0.1$. This
indicates that the radius of the region emitting Fe K$\alpha$ is smaller than half of the stellar radius $(<10^{12}~{\rm cm})$ 
and far more compact than the tail structure responsible for the absorption variability profile. 

Outside the eclipse, the equivalent width of the Fe K$\alpha$ line is of the order of 100 eV.
Following Matt (2002) \cite{Matt2002} and assuming a spherical transmission geometry, this corresponds to a column 
density of N$_{\rm H}\sim 2\cdot 10^{23}$ cm$^{-2}$. As this additional absorption is not observed,
the region emitting Fe K$\alpha$ must be partially ionized. 

 It is therefore very likely that the dense 
cocoon corresponds to the inner and ionized region of the hydrodynamical tail.
We thus added this partially ionized cocoon in our simulations, using a density
of $3\cdot 10^{11}~{\rm cm}^{-3}$ within a radius of $\sim 6\cdot10^{11}$ cm $\sim 3 \,{\rm R_{acc}}$.
The Fe K$\alpha$ emissivity map (fig. 2 right panel) was calculated by applying an illuminating radiation field $(\sim 1/r^{2})$
to the density distribution.
Figure 1 right panel displays the resulting simulated profile of the Fe K$\alpha$ equivalent width
together with the observed data. 
The green curve shows the variations in the Fe K$\alpha$ equivalent width expected from the wind density profile 
excluding the central cocoon, which
obviously could not reproduce the data.
The black curve accounts for the dense central cocoon. 
The exact profile of the eclipse is related to the size and density profile of the cocoon. 
No effort has been made to obtain an exact match to the data.

The majority of the Fe K$\alpha$ is formed in a region that is small 
enough to allow for pulsation of the Fe K$\alpha$ line.
We searched for such pulsations in our 
longest and almost uninterrupted observation.
Folded lightcurves were built in the energy bands 6.2 -- 6.7 keV, 2 -- 6.1 keV, and 6.8 -- 10 keV and 
resulted in a pulse fraction of $49\pm5$\%, $58\pm3$\%, and $57\pm 4$\%, respectively.
The ratio of the line flux to the continuum in the energy range  6.2 -- 6.7 keV   
is $\sim$0.25. Assuming that the line is not pulsed,    
we infer an Fe K$\alpha$ pulse fraction of $\sim$ 50\%, which is in  good agreement with the above measurement. 
Weak Fe K$\alpha$ pulsation can be explained if the cocoon is isotropic.

\begin{figure}[h]
  \centering
  \includegraphics[height=0.40\textwidth,width= 0.38\textwidth,angle=270]{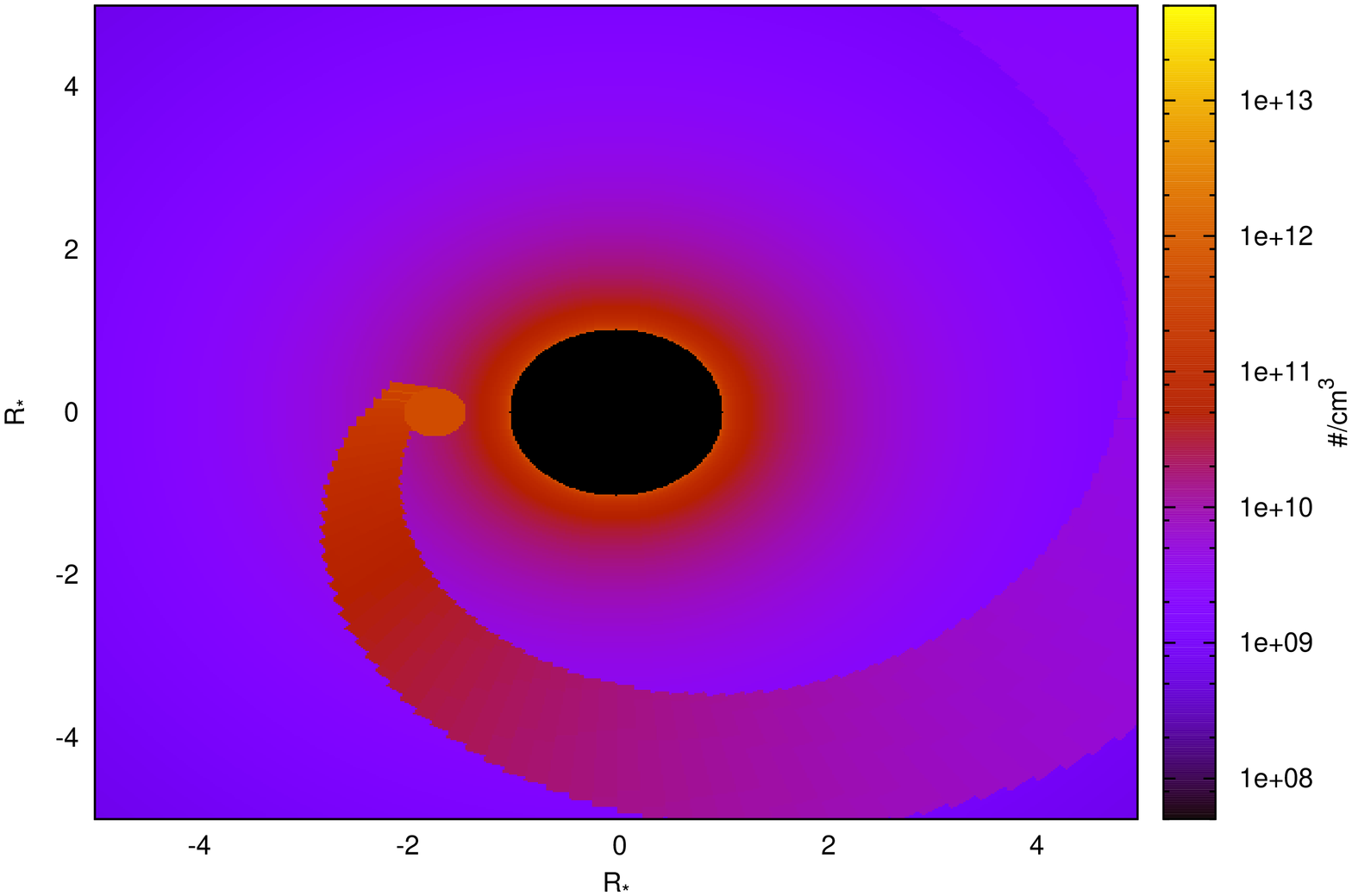} \hspace{1.5cm}
     \includegraphics[height=0.40\textwidth,width= 0.38\textwidth,angle=270]{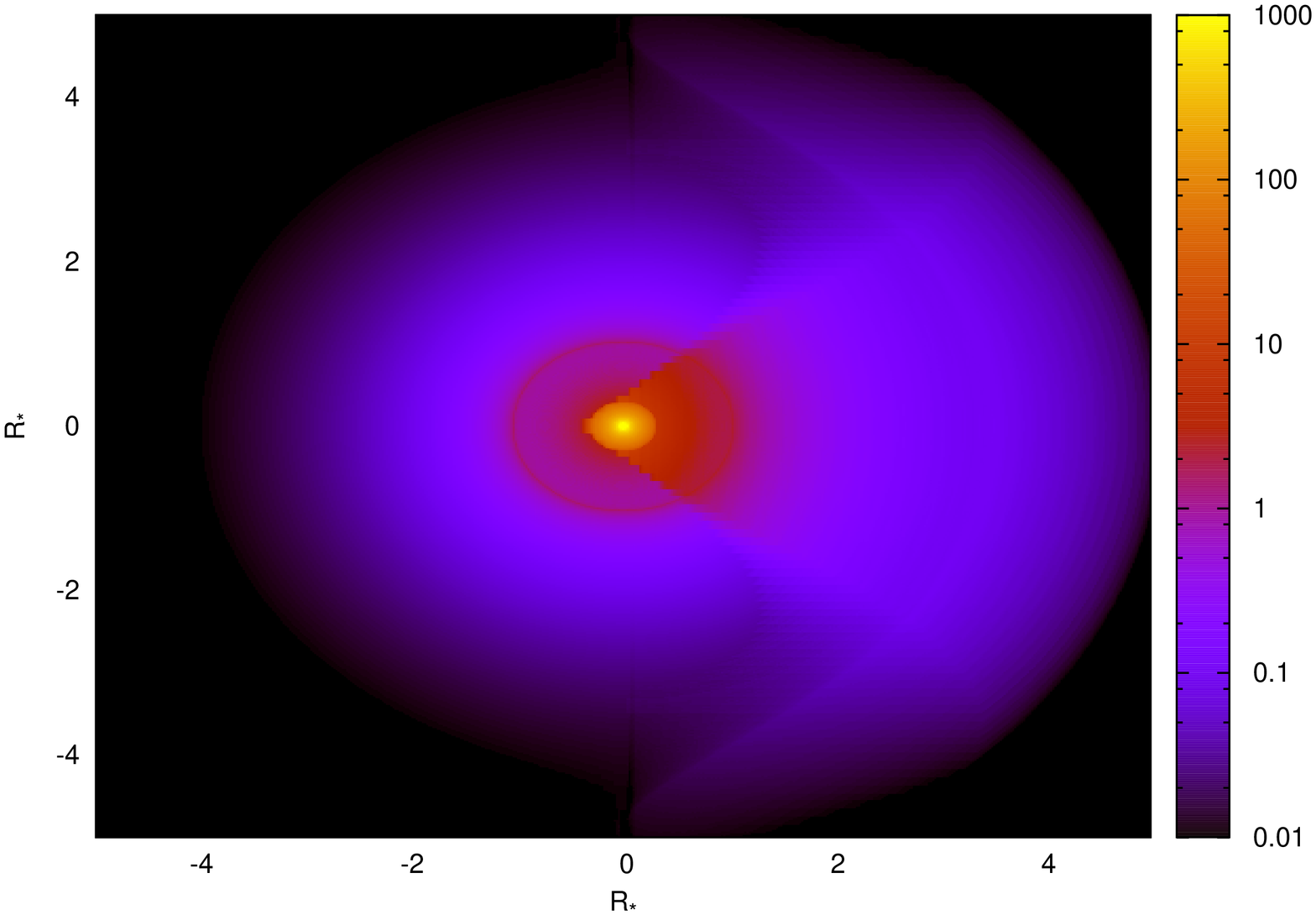}
    \caption{\emph{Left panel:} Number density distribution in the plane of the orbit including a smooth stellar wind and a tail-like perturbation. 
    The black disk at the center represents the supergiant companion. 
  \emph{Right panel:} Integrated iron K$\alpha$ emissivity (relative units) centered on the neutron star at phase $\phi=0.5$. The extended tail can be 
  seen on the right.
   }

  \label{fig:animals}
\end{figure}

\section{Summary}

The absorbing column density and the Fe K$\alpha$ emission line show remarkable variations.
The column density, always above $10^{23}$ cm$^{-2}$, increases towards $10^{24}$ cm$^{-2}$ close to 
the eclipse, as expected for a spherically symmetric wind. The wind velocity is unusually small
close to  $\upsilon_{\infty}=400$ km/s. An additional excess of absorption of $2\cdot10^{23}$ cm$^{-2}$ is 
observed for orbital phases $\phi>0.3$, which is found to represent a  hydrodynamical tail trailing behind the neutron star.

During the eclipse, the equivalent width of the Fe K$\alpha$ line drops by a factor $>10$ indicating
that most of the line is emitted in a cocoon surrounding the pulsar, with a size of a few accretion 
radii. This cocoon is ionized and corresponds to the inner region of the hydrodynamical tail

The parameters of the \igrj\, are very similar to these of Vela X-1, except for the smaller
wind velocity. We argue that the persistently large absorption column density is related to the 
hydrodynamical tail, which has been strengthened by the low wind velocity. The tail is a persistent structure 
dissolving on a timescale comparable to the orbital period.

Our interpretation can be tested using numerical hydrodynamical simulations and high resolution 
optical/infrared spectroscopy. If confirmed, half of the persistent sgHMXB may have 
stellar wind speeds several times lower than usually measured.

\bibliographystyle{plain}

\end{document}